\documentclass[aps,prl,twocolumn,groupedaddress]{revtex4}

\newcommand{\beq}{\begin{equation}}
\newcommand{\eeq}{\end{equation}}
\newcommand{\beqa}{\begin{eqnarray}}
\newcommand{\eeqa}{\end{eqnarray}}
\begin{document}

\title{Nuclei of Double-Charm Hyperons}

\author{B. Juli\'a-D\'{\i}az}
\email[]{Bruno.Julia@helsinki.fi}

\author{D. O. Riska}
\email[]{riska@pcu.helsinki.fi}
\affiliation{Helsinki Institute of Physics
and Department of Physical Sciences, POB 64,
00014 University of Helsinki, Finland}

\date{\today}
\begin{abstract}

The ground states of double-charm hyperons form a
spin $1/2$ isospin $1/2$ multiplet analogous to that
of nucleons. Their main strong interaction 
may be inferred directly from the corresponding 
nucleon-nucleon interaction by multiplication
of the interaction components by the appropriate fractional
difference between interaction strengths for pairs of light 
flavor quarks and pairs of triplets, e.g. nucleons, 
of light flavor quarks. By construction of the interaction
between the recently discovered double-charm hyperons
by this method from several realistic nucleon-nucleon
interaction models it is shown that double-charm hyperons
are likely to form bound (or possibly meta-stable)
states akin to the deuteron in the
spin triplet state. Double beauty baryons would form corresponding 
deeply bound states. Nucleons and double charm (beauty) hyperons 
will also form bound states. The 
existence of hypernuclei with double-charm and double-beauty
hyperons, which are stable against the strong decay, is very likely.

\end{abstract}


\maketitle

The SELEX collaboration has recently discovered several
double-charm hyperons, the lowest one of which is
the ground state multiplet with the $\Xi_{cc}^{++}$ at 
3.46 GeV/c$^2$ \cite{SELEX1, SELEX2}. 
The ground state multiplet $\Xi_{cc}^+$, $\Xi_{cc}^{++}$ forms
a spin $1/2$ isospin $1/2$ multiplet, with the valence quark
configuration $dcc$ and $ucc$~\cite{Richard}. 

The main color-neutral strong interaction between double-charm hyperons 
is due to their light flavor quark component.
This may be inferred from the nucleon-nucleon interaction, by multiplication
of the components of the nucleon-nucleon interaction by 
appropriate fractional coefficients, which relate the interaction
strength between pairs of light flavor quarks to that between
such triplets. The interaction between the charm
quark pairs in different hadrons is expected to be weaker  
than that between light flavor quarks in different 
hadrons, as the latter either arises from the short range
interaction that is mediated by the exchange of charmonia 
or from the color van der Waals interaction.
Because of the isospin $1/2$, the $\Xi_{cc}$ interactions
include a long range pion exchange component, in contrast to
the $\Lambda_c^+$, which has isospin 0. A priori hypernuclei
with double charm hyperon should therefore be more likely
than hypernuclei with single charm hyperons \cite{Dover}.

Although the strong interaction between double-charm hyperons is
weaker than that between nucleons, this is partially
compensated by their larger mass which 
weakens the repulsive effect of their
kinetic energy. The latter may be schematically illustrated
by a calculation of the binding energy of the deuteron-like
state, which is obtained by replacement of the nucleon mass by 
the mass of the $\Xi_{cc}$ (3.46 GeV/c$^2$), with the 
realistic nucleon-nucleon interaction models 
in Refs.~\cite{V18,NijmII,Paris}, 
all of which reproduce the experimental binding
energy of the deuteron and provide a quantitative
description of nucleon-nucleon scattering data. 
The corresponding binding energies are $-71$ MeV, 
$-75$ MeV and $-58$ MeV respectively
(the scatter between these values reflects the differences
in the detailed behavior of these interaction models at short
range, which is not constrained by nucleon-nucleon
scattering data).

In operator form the nucleon-nucleon interaction is given
in terms of rotational invariants of spin and isospin 
as well as momenta and angular momenta. The quark model
scaling factors between the matrix elements of the 
spin-isospin invariants for $\Xi_{cc}$ and nucleon states
may be derived from the quark
model matrix elements of light flavor quark operators~\cite{Close}:
\beqa
\langle \Xi_{cc}\vert 1 \vert \Xi_{cc}\rangle
&=&  {1\over 3} \langle N\vert 1 \vert N\rangle\, ,\nonumber\\
\langle \Xi_{cc}\vert \sum_q\sigma_a^q \vert \Xi_{cc}\rangle
&=&  -{1\over 3} \langle N\vert \sum_q\sigma_a^q \vert N\rangle\, ,\nonumber\\
\langle \Xi_{cc}\vert \sum_q\tau_a^q \vert \Xi_{cc}\rangle
&=&  \langle N\vert \sum_q\tau_a^q \vert N\rangle\, ,\nonumber\\
\langle \Xi_{cc}\vert \sum_q\sigma_a^q\tau_b^a \vert \Xi_{cc}\rangle
&=& -{1\over 5} \langle N\vert \sum_q\sigma_a^q \tau_b^q\vert N\rangle\, .
\label{scaling}
\eeqa
Here $N$ represents the nucleon.

The interaction between two double-charm hyperons (and that between 
double-charm hyperons and nucleons) that arises from
the interaction between the light flavor quarks
may be determined from any 
realistic nucleon-nucleon interaction model, that is given in
operator form. We here consider the models of 
Refs.~\cite{V18,NijmII,Paris} mentioned above. From these, after
dropping the (small) flavor symmetry breaking terms, which are 
inapplicable to hyperons (from those which include
such), the corresponding interactions between
double-charm hyperons may be derived by application of
the appropriate downscaling of the strengths of
the corresponding interaction
components. 

\begin{table}[h]
\begin{ruledtabular}
\begin{tabular}{l|cc}
Operator  &Argonne V18~\protect\cite{V18}  \\
\hline
\hline
Operator  & Scaling factor                    \\
\hline
1                             &$1/9$       \\
$\tau_i\!\cdot\!\tau_j$       & 1            \\
$\sigma_i\!\cdot\!\sigma_j$   &$1/9$            \\
$(\sigma_i\!\cdot\!\sigma_j)(\tau_i\!\cdot\!\tau_j)$  & $1/25$  \\
$S_{ij} $                       &  $1/9$    \\
$ S_{ij}(\tau_i\!\cdot\!\tau_j)$& $1/25$  \\
  $  {\bf L}\!\cdot\!{\bf S}$ & $-1/9$   \\
$ {\bf L}\!\cdot\!{\bf S}(\tau_i\!\cdot\!\tau_j)$& $-1/5$  \\
$  L^2$& $1/9$  \\
$L^2(\tau_i\!\cdot\!\tau_j)$& 1  \\
$L^2(\sigma_i\!\cdot\!\sigma_j)$& $1/9$    \\
$L^2(\sigma_i\!\cdot\!\sigma_j)(\tau_i\!\cdot\!\tau_j)$&$1/25$    \\
$({\bf L}\!\cdot\!{\bf S})^2$&  $1/9$   \\
$ ({\bf L}\!\cdot\!{\bf S})^2(\tau_i\!\cdot\!\tau_j)$ & 1   \\
\end{tabular}
\end{ruledtabular}
\caption{Quark scaling factors for the 
components of the interaction models of the form~\cite{V18,argonne}. 
\label{tab:sca}} 
\end{table}

In Table~\ref{tab:sca} the scaling factors for all 
the $NN$ operators that appear in the $NN$ potential 
models considered are listed. Where necessary, the 
larger scaling factor corresponding the operator 
has been taken. The main qualitative difference between the nucleon-nucleon 
interaction and that between double-charm hyperons is the 
weakening by the factor 9 of the central spin and isospin 
independent component, which in the case of
nucleons provides both the intermediate attractive
interaction and the strong short range repulsion. The fact
that the strengths at short range of the components of the
three nucleon-nucleon interaction models differ considerably
is of little numerical significance for the description of
nucleon-nucleon scattering observables at low energy, because
of the strong short range repulsion. The weakening of that
repulsive interaction component in the double-charm
hyperon interaction increases the significance of the
differences in the short range interaction components, which is
revealed as large differences in the predicted binding
energies.

For the deuteron the main cause of binding is the strong isospin
dependent tensor force, the long range part of which 
is due to pion exchange. Since this is weaker by a factor
25 for double-charm hyperons, it follows that its 
role is significantly smaller for double charm hyperons.
Below we show, on the basis of the quark model scaling
factors, that two of the three rescaled nucleon-nucleon interaction
models nevertheless do imply that double-charm hyperons
form deuteron-like bound states, with binding energies
in the range $87-457$ MeV.  

The binding energies were calculated by diagonalization 
of the Hamiltonian in a discretized Gauss-Legendre mesh in 
momentum space. The bound state equation can be written as, 
\beq
\sum_{l,l'} \int  dq' q'^2 
\bigg[ {q^2 \over 2 \mu} 
{\delta_{ll'} \delta(q-q')\over q'^2} + V_{ll'}(q,q') \bigg] \phi_{l'}(q')
=E \phi_l(q) \, .
\eeq
Here $l$ and $l'$ are orbital angular momenta, $\mu$ is the 
reduced mass of the system and $V_{ll'}(q,q')$ is the 
partial wave decomposed potential ($<qlsj|V|q'l'sj>$). 
For the purpose of the present 
investigation a mesh of 80 Gauss-Legendre points was 
sufficient for stable results. As a test of 
the numerical method the deuteron binding, $-2.2$ MeV,
was calculated with all the models, AV18, Nijmegen II and 
Paris potential. 

In Table~\ref{table1} the calculated binding energies 
obtained for the deuteron-like states of double-charm hyperons
are given.
The scatter between the calculated values provides
an estimate of the theoretical uncertainty that 
derives from the different short range behavior of the
nucleon-nucleon interaction models. 

It is natural to assume that the experimentally
discovered double-charm hyperon state $\Xi_{cc}^{+}(3520)$ 
represents the double-charm spin $3/2$ state, which is analogous
to the $\Delta(1232)$ \cite{Richard,sco}. The 
60 MeV splitting between
this state and the  $\Xi_{cc}^{++}(3460)$ is in line
with quark model estimates~\cite{Coester}, although
slightly smaller than numerical lattice method
based estimates~\cite{Flynn}. As the coupling to this
state can only increase the calculated binding energy,
the binding energies in Table \ref{tab:sca} may
be viewed as lower bounds.

Two-baryon states formed of double-charm hyperons can
in principle couple to states with a single charm and
a triple-charm $\Omega_{CCC}$ by the quark rearrangement 
interaction. If the latter states have
lower energy the former are metastable rather than bound.
This depends on the size of the binding energy as 
compared to the mass difference:
\beq
\Delta^c \equiv M_{ccc}+M_{cll} - 2 M_{ccl} 
\label{ine}
\, .
\eeq
Here $l$ represents a light quark.
For many non-relativistic quark models the 
inequality $\Delta^c <0$ holds~\cite{jmrscales}. 
Corresponding numerical estimates suggest that
$\Delta \approx [130 - 158]$ MeV \cite{silvestre}.
Adoption of those values imply that the double charm
hyperons form bound states with
the AV18 potential, but only metastable states with the 
Nijm II potential. 

\begin{table}
\begin{ruledtabular}
\begin{tabular}{ll}
\multicolumn{2}{c}{Double-Charm hyperons}\\
\hline
Potential        & Binding Energy (MeV) \\
AV18             &  $-$457 ($-$28)   \\
Paris            &  $-$              \\
Nijm II          &  $-$87            \\
\hline
\hline
\multicolumn{2}{c}{Double-Beauty double-charm hyperons}\\
\hline
AV18             & $-603$ ($-183$)    \\
Paris            & $-0$               \\
Nijm II          & $-102$             \\
\hline
\hline
\multicolumn{2}{c}{Double-Beauty hyperons}\\
\hline
AV18             &  $-$782 ($-$439)   \\
Paris            &  $-$2        \\
Nijm II          &  $-$123 ($-$20)     \\
\end{tabular}
\end{ruledtabular}
\caption{Binding energies for the $\Xi_{cc}^{++}-\Xi_{cc}^+$ and 
$\Xi_{bb}^0-\Xi_{bb}^-$ systems obtained with Argonne 
V18~\protect\cite{V18}, AV18, 
Nijmegen II~\protect\cite{NijmII} and Paris~\protect\cite{Paris} 
potentials. For the double-beauty baryons the mass values: 
$M_{bbu}=M_{bbd}=9000$ MeV were used. The value in brackets corresponds 
to a second bound state.\label{table1}}
\end{table}

To further explore the theoretical uncertainty of the
calculated binding energies these have also been
calculated using the class ``AVn'' of systematically
simplified versions of the V18 \cite{V18} interaction
models \cite{argonne}. These results are shown in
Table~\ref{table3}. The number of each potential designates the 
number of operator structures present,  ordered as in the 
first column of Table~\ref{tab:sca}. These potential
models are reprojections of the full AV18 potential,
of which all but AV1' and AVX' reproduce the 
deuteron binding energy.

The results show a considerable spread in the calculated binding 
energies. No bound state is found with
the AV8' potential, which already 
includes tensor and spin orbit forces. 
With the simpler AV2', AVX', AV4' 
and AV6' models bound states appear. 
Consequently nucleon-nucleon interaction models, 
which give the correct deuteron binding energy, with (AV6') and without 
tensor or spin-orbit forces, predict a bound state 
in the $\Xi_{cc}^{++}-\Xi_{cc}^+$ system in the spin triplet
state. The case of the AV2' potential, for which the quark model 
scaling is simply a factor 1/9 of the {\bf 1} part, and 
which predicts a bound state of 26 MeV, is particularly
telling. The binding energies 
obtained with the AVn' class of potentials are 
considerably smaller than the one obtained with the full AV18 
potential. In particular, with the numerical
estimates in Ref~\cite{jmrscales} 
based on the mass inequalities, see Eq.~(\ref{ine}), 
these bound states would be metastable against decay to
the single-charm -- triple-charm hyperon state.

\begin{table}

\begin{ruledtabular}
\begin{tabular}{l|c|c|c|c}
AVn'            & $\Xi_{cc} \Xi_{cc}$ & $\Xi_{cc} \Xi_{bb}$   
&$ \Xi_{bb} \Xi_{bb}$ & deuteron \\
\hline 
Potential       &\multicolumn{4}{c}{Binding Energies (MeV)} \\
\hline
AV8'            & $-$          & $-0$           & $-3$           & $-2.2$  \\
AV6'            & $-1$         & $-5$           & $-15$          & $-2.2$  \\
AV4'            & $-30$        & $-41$          & $-58$ ($-4$)     & $-2.2$  \\
AVX'            & $-15$        & $-22$          & $-33$          & $-0.4$  \\
AV2'            & $-26$        &$-48$           & $-88$ ($-4$)     & $-2.2$  \\
AV1'            & $-$          & $-$            & $ -$           & $-0.4$  \\
\end{tabular}
\end{ruledtabular}
\caption{Estimated binding energies for the pairs
$\Xi_{cc}-\Xi_{cc}$, $\Xi_{cc}-\Xi_{bb}$ and $\Xi_{bb}-\Xi_{bb}$ of
double-heavy together with the deuteron as obtained with the class of  
Argonne potentials in Ref.~\cite{argonne}.\label{table3}}
\end{table}

The origin of the large binding energy given by the AV18 interaction
model may be traced to its large isospin dependent squared spin-orbit
interaction, which acts in the $D-$state. The relative
significance of this interaction component derives from
the large quark model scaling factor (Table 1). The general
reason for the large spread in the calculated binding energies
obtained with the different interaction models is the fact that
their components differ notably at ranges less than 0.6 fm.
These differences are of little significance for 
nucleon-nucleon scattering observables at low energy, because
of the very strong short range repulsion. The reduction of the
strength of the repulsive central interaction by a factor 9
enhances the role of the short range behavior of the state
dependent components of the interaction in the case of
double-charm hyperons.

Double beauty-hyperons analogous to, but heavier than,
the double-charm hyperons are expected to exist in the mass
range above twice the B-meson mass. The interaction
between the ground state multiplets of double-beauty
hyperons with the quark configurations 
$(bbu)$ and $(bbd)$ is expected to be very similar to that
of the double charm hyperons, as their main interaction is
that between their light flavor quarks. Their binding 
energy will however be much larger than that of double-charm
hyperons in view of their larger mass. This is illustrated
in Table~\ref{table1} where the binding energy of two
double-beauty hyperons is estimated by the same method as
used above for double charm hyperons with the assumption
that their mass is: 
$M_{bbu}=M_{bbd}\approx 2 m_b \approx 2 \times 4500 $ MeV.
In Table~\ref{table3} the binding energies obtained 
with the AVn' potentials are listed for comparison as well.
The majority of these 
potential models yield a bound state, in some cases two, with
binding energies  
up to few hundred MeV. 

The possibility of metastability is in this case 
bounded by a similar inequality,  
as in the case of the charmed hyperon case: 
$\Delta^b \equiv M_{bbb}+M_{bll} - 2 M_{bbl} <0$, where 
its estimated value ranges $[348 - 372]$ 
MeV~\cite{jmrscales,silvestre}. As in the case of the 
double-charm hyperons the result 
obtained with the AV18 potential is a bound state,
while that obtained with the
Nijm II potential is a metastable state.

Finally we have also explored the possibility of 
deuteron-like bound states
of nucleons and double-heavy hyperons:
$N-\Xi_{cc}$ and $N-\Xi_{bb}$. 
Such bound states were 
found with the AV18 and the Nijm II potential models. The AV18 
potential gives bound states at $-388$ MeV and $-494$ MeV 
for the $N-\Xi_{cc}$ and $N-\Xi_{bb}$ systems respectively.
The Nijm II potential gives bound states at $-35$ MeV and $-76$ MeV 
for the $N-\Xi_{cc}$ and $N-\Xi_{bb}$ systems respectively. 

In summary it has been shown that bound states of the 
recently discovered double-charmed hyperons are likely.  
Their binding energies were estimated by construction 
of their main strong interaction by the quark-model scaling 
relations from several realistic phase-equivalent
nucleon-nucleon interaction models.
The $\Xi_{cc}$ hyperons are likely to be bound in the
spin triplet state by more than 30 MeV. The $\Xi_{bb}$
hyperons are expected to be bound by more than 50 MeV
in the triplet state, while pairs of $\Xi_{cc}$ and
$\Xi_{bb}$ baryons are expected to be bound with binding
energies of that order. The qualitative nature of these
estimates reflects the uncertain short range behavior
of the extant realistic nucleon-nucleon interaction models.

\begin{acknowledgments}
The authors want to thank Frank Fr\"omel for pointing out 
an algebraic error in the original version of this
manuscript. B.J.-D thanks 
J. M. Richard for stimulating discussions and
the European Euridice network for support 
(HPRN-CT-2002-00311). Research supported in part 
by the Academy of Finland through grant 54038.

\end{acknowledgments}

%\newpage

\end{document}